\documentclass[sigconf,natbib=true,anonymous=false]{acmart}

\usepackage{epigraph}

\AtBeginDocument{%
  }

\setcopyright{acmlicensed}
\copyrightyear{2018}
\acmYear{2018}
\acmDOI{XXXXXXX.XXXXXXX}
\acmConference[Conference acronym 'XX]{Make sure to enter the correct
  conference title from your rights confirmation email}{June 03--05,
  2018}{Woodstock, NY}
\acmISBN{978-1-4503-XXXX-X/2018/06}

\begin{document}

\title{InterDeepResearch: Enabling Human-Agent Collaborative Information Seeking through Interactive Deep Research}
\author{Bo Pan}
\affiliation{%
  \institution{State Key Lab of CAD\&CG \\ Zhejiang University}
  \city{Hangzhou}
  \country{China}
}
\email{bopan@zju.edu.cn}

\author{Lunke Pan}
\affiliation{%
  \institution{State Key Lab of CAD\&CG \\ Zhejiang University}
  \city{Hangzhou}
  \country{China}
}
\email{3230102935@zju.edu.cn}

\author{Yitao Zhou}
\affiliation{%
  \institution{State Key Lab of CAD\&CG \\ Zhejiang University}
    \city{Hangzhou}
  \country{China}
}
\email{3230102935@zju.edu.cn}

\author{Qi Jiang}
\affiliation{%
  \institution{State Key Lab of CAD\&CG \\ Zhejiang University}
    \city{Hangzhou}
  \country{China}
}
\email{3230102911@zju.edu.cn}

\author{Zhen Wen}
\affiliation{%
  \institution{State Key Lab of CAD\&CG \\ Zhejiang University}
    \city{Hangzhou}
  \country{China}
}
\email{wenzhen@zju.edu.cn}

\author{Minfeng Zhu}
\affiliation{%
  \institution{Zhejiang University}
    \city{Hangzhou}
  \country{China}
}
\email{minfeng_zhu@zju.edu.cn}

\author{Wei Chen}
\affiliation{%
  \institution{State Key Lab of CAD\&CG \\ Zhejiang University}
    \city{Hangzhou}
  \country{China}
}
\email{chenvis@zju.edu.cn}
\renewcommand{\shortauthors}{Trovato et al.}

\begin{abstract}
  Deep research systems powered by LLM agents have transformed complex information seeking by automating the iterative retrieval, filtering, and synthesis of insights from massive-scale web sources. However, existing systems predominantly follow an autonomous ``query-to-report'' paradigm, limiting users to a passive role and failing to integrate their personal insights, contextual knowledge, and evolving research intents. This paper addresses the lack of human-in-the-loop collaboration in the agentic research process. Through a formative study, we identify that current systems hinder effective human-agent collaboration in terms of process observability, real-time steerability, and context navigation efficiency. Informed by these findings, we propose \textit{InterDeepResearch}, an interactive deep research system backed by a dedicated research context management framework. The framework organizes research context into a hierarchical architecture with three levels (information, actions, and sessions), enabling dynamic context reduction to prevent LLM context exhaustion and cross-action backtracing for evidence provenance. Built upon this framework, the system interface integrates three coordinated views for visual sensemaking, and dedicated interaction mechanisms for interactive research context navigation. Evaluation on the Xbench-DeepSearch-v1 and Seal-0 benchmarks shows that \textit{InterDeepResearch} achieves competitive performance compared to state-of-the-art deep research systems, while a formal user study demonstrates its effectiveness in supporting human-agent collaborative information seeking. Project page with system demo: \url{https://github.com/bopan3/InterDeepResearch}.

\end{abstract}

\ccsdesc[500]{Information systems~Users and interactive retrieval}

\keywords{Deep Research, Agent, Human-Agent Collaboration}

\received{20 February 2007}
\received[revised]{12 March 2009}
\received[accepted]{5 June 2009}

\maketitle

\section{Introduction}
\sloppy
Recently, deep research systems have proliferated across both academia \cite{xu2025comprehensive, zhang2025deep} and industry \cite{openai_deep_research,gemini_deep_research,kimi_researcher}. By equipping LLM-based agents with information search tools and enabling them to iteratively invoke these tools and analyze search results, these systems can address complex research goals posed by users, demonstrating substantial economic potential across diverse application scenarios, such as literature reviews \cite{alphaxiv_explore,citely_source_finder}, market analysis \cite{lev8_2026, xbench_marketing} and domain-specific investigations \cite{openevidence,chen2025medbrowsecomp,jin2025finsight,hu2025finsearchcomp}.

Despite these advances, current deep research systems are predominantly designed and optimized for a single-turn query-to-report paradigm, limiting users to a passive role without any opportunity to participate in the research process. However, a substantial body of research in information retrieval has established that many information seeking tasks benefit significantly from human-in-the-loop collaboration, particularly when users need to contribute personal insights, contextual knowledge, and evolving research intents that cannot be fully captured in initial queries \cite{bates1990interface,marchionini1995information,white2016interactions}. This need for user involvement manifests in two ways during deep research. At a high level, users may need to adjust research strategies based on emerging insights. For instance, upon observing preliminary findings that reveal an unexpected but promising research direction, the user might recognize the value of pivoting the investigation toward this new angle. At a low level, users may need to intervene on critical execution details, such as correcting the agent when it relies on incorrect assumption or specifying precise search keyword that require contextual knowledge. Yet existing deep research systems provide limited support for users to monitor the ongoing research process or make interventions, forcing users to either passively accept whatever the agent produces or abandon the entire research session and restart from scratch with a revised query.

This work addresses this gap by exploring how to build an interactive deep research system that supports human-agent collaborative information seeking. To systematically understand users' core challenges and needs when attempting to collaborate with deep research agents, we first conducted a formative study with frequent users of existing deep research systems. Through this study, we identify three key challenges that hinder effective collaboration: (1) unclear presentation of the agent's research process, which prevents users from developing accurate mental models of ongoing research activities; (2) limited support for timely steering of research direction, which deprives users of the opportunity to promptly correct deviations or adjust research strategies on the fly; and (3) inefficient navigation of massive research context, making it cumbersome for users to trace conclusions back to supporting evidence or locate specific information encountered during research.

Informed by these findings, we distilled design goals and proposed \textit{InterDeepResearch}, an interactive deep research system specifically designed for human-agent collaborative information seeking. To achieve the design goals, we introduce a novel research context management framework that considers the information processing characteristics of both humans and agents. The framework proposes a hierarchical research context architecture that organizes the research process across three levels: research information (the atomic units of accumulated information), research actions (the interrelated operations that percieve and produce research information), and research sessions (macro-level groupings of research actions). This hierarchical architecture enables both comprehensive visibility into research details and efficient macro-level progress tracking, while also supporting context reduction to prevent context exhaustion during long-horizon research, and context backtrace to help trace conclusions to their supporting evidence embedded in massive research context.

The interface of \textit{InterDeepResearch} featuring three coordinated views to support clear comprehension of agent research processes (see Figure \ref{fig:system_interface}): a linear chat-style view for research session marking and narrative research action flow, a graph-based canvas-style view for research action dependency, and a card-style view for detailed research information. Users can interrupt the agent at any point and  steer the research direction flexiblely like chatting with a research partner. To help users efficiently navigate the research context, we provide cross-view linkage that allows users to focus on any research action and instantly retrieve corresponding information across all views, as well as cross-action backtrace that enables users to easily trace conclusions back to their supporting evidence.

We evaluated \textit{InterDeepResearch} through a technical evaluation and a formal user study. Our results show that InterDeepResearch achieves competitive performance on existing deep research benchmarks and effectively supports human-agent collaborative information seeking. We summarize our contribution as follows:

\begin{itemize}
\item \textit{InterDeepResearch}, the first interactive deep research system for human-agent collaborative information seeking.
\item A formative study that identifies key challenges and design goals for human-agent collaborative information seeking.
\item A novel research context management framework considering the information processing characteristics of both humans and agents during collaborative information seeking.
\item A technical evaluation and a formal user study demonstrating the effectiveness of our system in supporting human-agent collaborative information seeking.
\end{itemize}

\section{Related Work}
\subsection{Deep Research Systems}
Deep research is an emerging information-seeking application that leverages LLM-based agents to conduct long-horizon information search, collection and synthesis. Traditional information retrieval systems typically rely on hard-coded retrieval workflows, while Retrieval-Augmented Generation (RAG) \cite{izacard2022atlas, gao2023retrieval, sharma2025retrieval} leverages LLMs' general intelligence to process intermediate information but still maintains rigid workflows. Deep research systems \cite{xu2025comprehensive, zhang2025deep}, in contrast, fully exploit the autonomy of LLM-based agents, enabling them to proactively plan, invoke information retrieval tools, and process information like human researchers, thereby accomplishing more complex and long-horizon information seeking tasks. 

Many methods have been proposed to optimize the information seeking capabilities of deep research systems, which can be categorized into three major directions: architecture-based, supervision-based, and reward-based approaches. Architecture-based methods improve system performance through carefully designed agent architectures. For example, WebWalker \cite{wu2025webwalker} designs an explore-critic paradigm to balance research breadth and accuracy. Anthropic proposes a leader-searcher multi-agent architecture to maximize investigation efficiency \cite{anthropic2025multiagent}. Google's test-time diffusion framework \cite{han2025deep} models the agent's research process as a continuous information denoising process to tackle performance plateaus. Supervision-based methods optimize models by collecting high-quality deep research trajectories as training data. SimpleDeepSearcher \cite{sun2025simpledeepsearcher} proposes a real web-based data synthesis framework that simulates realistic user search behaviors and employs multi-criteria curation strategies to ensure data quality. WebSailor \cite{li2025websailor} generates high-uncertainty training tasks through structured sampling from knowledge graphs and uses open-source reasoning models to generate solution traces while reconstructing concise reasoning for supervision. Reward-based methods employ reinforcement learning to optimize agent behavior. For instance, DeepResearcher \cite{zheng2025deepresearcher} trains multi-agent systems using group relative policy optimization with rewards based on format correctness and answer accuracy. InForage \cite{qian2025scent} augments outcome-based rewards with information gain and efficiency penalties to discourage redundant reasoning.

Despite these advances, prevailing deep research systems are designed around one-way, single-query interactions that confine users to passive roles without opportunities for engagement or influence. Our work introduces an interactive deep research system enabling active user participation, making an early attempt to examine both the challenges and potential of human-agent collaborative information seeking.

\subsection{Interface for Information Seeking}
User interfaces play an important role in designing effective systems for information seeking \cite{bates1990interface,white2016interactions}. The field spans simple keyword-based query boxes to sophisticated interactive systems supporting complex exploratory behaviors \cite{hearst2009search, white2009exploratory, Marchionini2006exploratory}. Early research on interactive information retrieval interfaces introduced a set of methods to facilitate query refinement, such as explicit relevance feedback \cite{ruthven2003survey} and faceted search \cite{Facetedsearch}. Visual interfaces like SearchLens \cite{chang2019searchlens} and SenseMap \cite{SenseMap} leverage spatial organizations to support sensemaking as information accumulates, while some works extend the dimensionality of the interface to further augment users' cognitive abilities \cite{north2012analyst, lisle2020evaluating, Yang2025LitforagerEM}. For multi-user collaborative information seeking, systems like SearchTogether \cite{Morris2007SearchTogether} and Coagmento \cite{Shah2009CoagmentoAC} provide shared workspaces with awareness indicators and synchronized annotations to support team-based exploratory search. Recently, the integration of LLMs and traditional search interfaces has attracted researchers' attention \cite{Zerhoudi2025SearchLab, Zhang2024GenSERP, xiong2024searchenginservicesmeet}, and various interfaces tailored for LLM-based information exploration scenarios have also been proposed \cite{suh2023sensecape, jiang2023graphologue, suh2024luminate}.

Extending this line of work, our work explores how to build effective interfaces that support collaborative information seeking between humans and LLM-based agents.

\section{Formative Study}

We conducted a formative study to systematically understand users' core challenges and needs when collaborating with deep research agents for information seeking. The findings from this study provided guidance for the design of \textit{InterDeepResearch}.

\subsection{Participants and Procedure}
To identify corresponding needs insufficiently addressed in existing deep research systems and inform the design of our system, we conducted a formative study with 8 frequent users (P1-P8) who extensively utilize existing deep research systems. Among the participants, two had experience developing deep research systems, four regularly used existing deep research systems in their work, and five frequently employed such systems in their daily lives.

For each participant, we first asked them to report their usage patterns in work and daily life, demonstrate the deep research systems they typically use, and show how they typically use these systems to solve their information-seeking tasks. We then asked participants to propose an information-seeking task that they would want to actively participate in, and enter the task description into their preferred deep research system to execute it. For deep research systems that stream and display information about the research process in real-time, we asked participants to observe this displayed information and use ``think aloud'' approach to indicate when they would want to intervene in the research process and how they would want to intervene. After the deep research system completed the research and generated a final report, we asked participants to reflect on how they would like to intervene in or lead the research process if they could collaborate with the agent, and to identify current challenges in engaging the collaborative research process. We also encouraged participants to share features they believed would help them collaborate better with the agent, and maintained regular contact with them throughout the subsequent iterative prototyping process to collect their feedback.


\subsection{Findings}
All participants believed that enabling users to more actively participate in the deep research process would be useful and sometimes highly desirable. We summarize the key challenges for users to intervene in or lead the human-agent collaborative information seeking that emerged from the study as follows.

\textbf{C1. Unclear presentation of agent's research process.}
Although most commercial deep research systems provide visibility into the agent's research process, these displays are typically simplified and incomplete. This makes it difficult for users to see specific research details they care about, track overall progress, or understand dependencies between research actions. For example, one user of OpenAI Deep Research noted, ``I can see the system showing that it's searching for something about a topic one moment, and reading a webpage the next, but the information displayed is very limited. I don't know what specific content it's searching for or reading'' (P2). Multiple participants mentioned difficulty ``understanding the current overall state of the research'' (P2, P3, P8). Additionally, some participants noted ``difficulty understanding the dependencies between various agent actions'' (P1) and expressed a desire to ``more clearly see how each step contributes to answering the question'' (P8). This insufficient observability made it difficult for participants to identify potential issues or opportunities for intervention during the research process.

\textbf{C2. Limited support for timely steering of research direction.}
All participants expressed a desire for mechanisms to steer the agent's research direction in real-time. However, most existing deep research systems confine users to a narrow interaction model of initial query submission and final report reception. P7 stated that ``sometimes I notice the agent's research direction deviating from my original expectations, but I have no way to correct it—I can only watch helplessly as it continues down that path'' (P7). P2 mentioned that they sometimes ``want to tell the agent to conduct more divergent breadth-first exploration rather than getting stuck in certain details, or after discovering some key information during the research process, to have the agent focus on deep investigation following that lead'' (P2). The lack of intervention mechanisms meant that participants had to wait for the entire research process to complete before they could provide feedback through a follow-up query, which would then trigger another lengthy research cycle. Moreover, it is hard to access the research context from when users wanted to intervene in the new research cycle. Participants suggested features such as the ability to pause the research, adjust priorities on different research directions, or inject new requirements mid-process to steer the research direction more flexibly.

\textbf{C3. Inefficient navigation of massive research context.}
Existing deep research systems compile all the complex, interconnected information encountered during the research process into a single lengthy textual report with citations. Participants reported difficulty efficiently navigating the report and raw information sources to locate specific information they needed during iterative analysis. P5 told us that ``often I want to trace the sources of certain conclusions in the report, but finding the information sources that support the corresponding conclusions and then locating the relevant supporting evidence within those sources is very cumbersome'' (P5). Additionally, some participants mentioned that establishing connections between the results in the final report and the corresponding research process is also troublesome, and expressed a desire for better features to support such mapping (P2, P7).

\subsection{Design Goals}
Based on the findings from our formative study and an iterative prototyping process, we distilled three core design goals for building a user-centered interactive deep research system.

\textbf{G1. Support clear comprehension of agent research processes.}
The system should provide comprehensive visibility into the agent's research process, helping users understand research progress at the macro level while supporting on-demand access to details of individual research actions and their dependencies at the micro level.

\textbf{G2. Enable flexible steering of agent research direction.}
The system should support mechanisms for users to timely intervene and guide the agent's research direction. Users should be able to pause research, inject new requirements, adjust strategies, and redirect focus based on emerging insights without restarting the entire research cycle.

\textbf{G3. Facilitate efficient navigation of massive research context.}
The system should provide interaction mechanisms for users to efficiently navigate research context, quickly locate information related to specific results, and trace the provenance of conclusions back to supporting evidence.

\section{Research Context Management}
\label{section:Research Context Management}
\begin{figure*}[t]
  \centering
  \includegraphics[width=\textwidth]{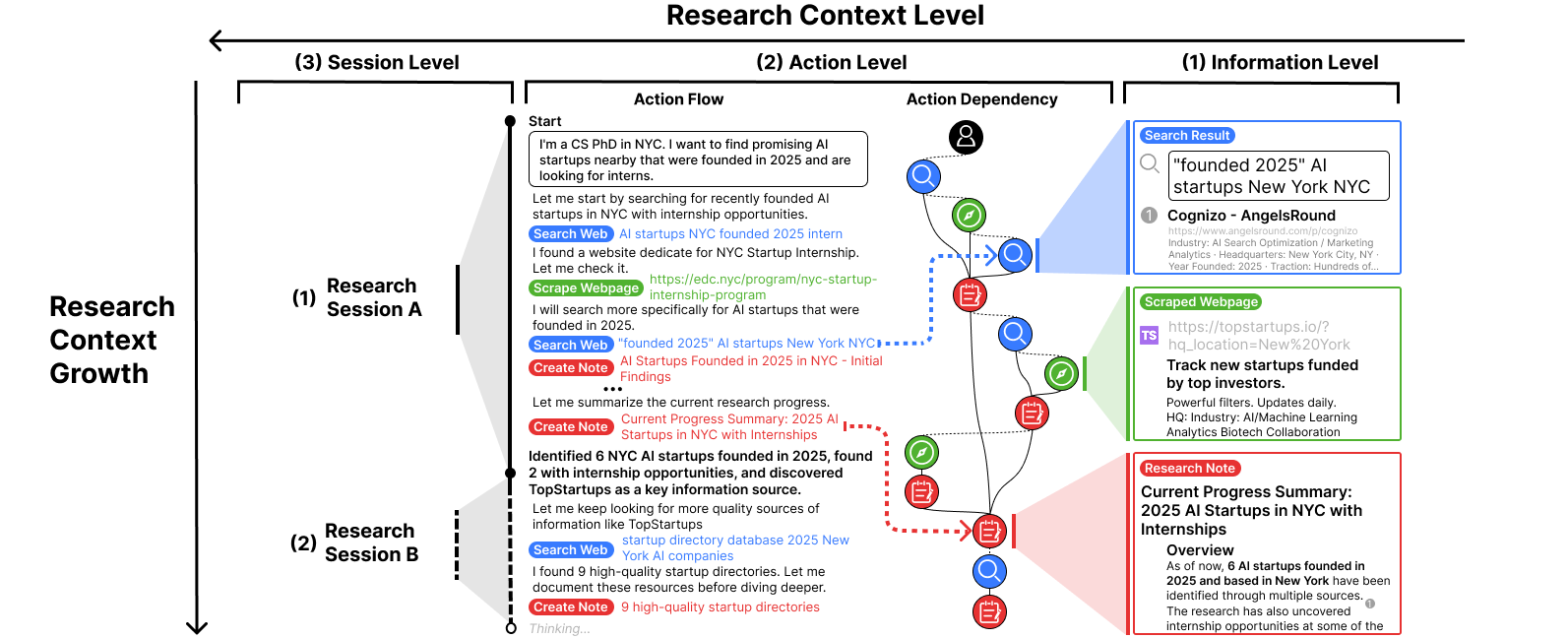}
  \vspace{-10pt}
  \caption{The hierarchical research context architecture across three levels: information, actions, and sessions.}
  \Description{A diagram showing the three-level hierarchical architecture of the research context management system. The bottom level shows research information nodes, the middle level displays research actions with their dependencies, and the top level illustrates research sessions grouping multiple actions together.}
  \label{fig:context_management}
  \vspace{-5pt}
\end{figure*}

Building an effective interactive deep research system requires a well-matched research context management mechanism. In this section, we introduce a novel research context management framework specifically designed for human-agent collaborative information seeking, based on our design goals and the information processing characteristics of both humans and agents.

\subsection{Hierarchical Research Context Architecture}
\label{section:Hierarchical Research Context Architecture}
To enable users to easily understand and grasp research progress at both macro and micro levels, we propose a hierarchical research context architecture, which is elaborated as follows. 

\subsubsection{Level1: Research Information}
At the most concrete level, the entire research process is regarded as a continuous information accumulation process which can be represented as:
\begin{equation}
    \mathcal{I}_{t+1} = \mathcal{I}_{t} \cup \{i_{t+1}\}
\end{equation}

where $\mathcal{I}_{t}$ represents the accumulated information set at step $t$, and $i_{t+1}$ represents the newly acquired information unit.

The information unit $i$ can be classified into 4 general types: 
\begin{itemize}
\item \textbf{User Information}: The information provided by the user. E.g., a research request with background information.
\item \textbf{Search Information}: The search result of search engine for external source information. E.g., a Google search result (see the blue information block in the right column of Figure \ref{fig:context_management}).
\item \textbf{Source Information}: The raw external source information. E.g., a scraped webpage (see the green information block in in the right column of Figure \ref{fig:context_management}).
\item \textbf{Processed Information}: The information generated by agent by processing other information. E.g., a generated summary based on several previous information (see the red information block in the right column of Figure \ref{fig:context_management}).
\end{itemize}

\subsubsection{Level2: Research Action}
Directly presenting all information to users would overwhelm them with massive amounts of concrete details, failing to provide a clear understanding of the research process. What helps users efficiently comprehend the research process is understanding why information was generated and the dependencies between different pieces of information. Therefore, we introduce the concept of research action, where each action is executed by either the agent or the user based on the prior research context. Different types of research actions have different parameters and produce different types of information.

Formally, the research process can be viewed as a continuous sequence of research actions:
\begin{equation}
a_{t+1} = f(\mathcal{A}_{t}, \mathcal{I}_{t}), \quad f \in \{\text{Agent}, \text{User}\}
\end{equation}

where $\mathcal{A}{t} = {a_1, ..., a_t}$ represents the history of executed research actions up to step $t$, $\mathcal{I}{t}$ represents the accumulated information at step $t$, and the next action $a_{t+1}$ is determined by the decision-making function $f$ (which can be either the agent or the user).

We classify research actions into five basic types:
\begin{itemize}
\item \textbf{User Information Action}: The action that generates \textbf{User Information}. E.g., the user can execute the ``User Message'' action (with the information to be sent as parameter) to send user message to the agent, or execute the ``User Interrupt'' action to interrupt the agent.
\item \textbf{Search Information Action}: The action that generates \textbf{Search Result Information} (see the blue nodes in Figure \ref{fig:context_management}). E.g., the agent can execute the ``Web Search'' action (with search terms as parameters) to produce \textbf{Search Information}.
\item \textbf{Source Information Action}: The action that generates \textbf{Source Information} (see the green nodes in Figure \ref{fig:context_management}). E.g., the agent can execute the ``Scrape Webpage'' action (with target webpage URL as parameter) to produce \textbf{Source Information}.
\item \textbf{Processed Information Action}: The action that generates \textbf{Processed Information} (see the red nodes in Figure \ref{fig:context_management}). E.g., the agent can execute the ``Create Note'' action (with input source references and processing requirements as parameters) to produce \textbf{Processed Information}.
\item \textbf{Administrative Action}: The administrative action that does not generate any information. E.g., the agent can execute the ``Finish'' action to stop.
\end{itemize}

We provide two complementary ways to organize research actions to facilitate the comprehension of the research process across different scenarios. For scenarios where the user/agent needs to follow the chronological progression, we provide a linear \textbf{Action Flow} (shown in the middle left column of Figure \ref{fig:context_management}), where the agent provides brief explanations before executing each action about what just happened and what it is about to do, creating a concise narrative of the research evolution. For scenarios where the user/agent needs to check the action dependency and information provenance, we provide a \textbf{Action Dependency Graph} (shown in the middle right column of Figure \ref{fig:context_management}), where the dependency between different actions is explicitly recorded.


\subsubsection{Level3: Research Session}
\label{Section:Level3: Research Session}
To enable users to understand the research process at a more macro scale, we introduce the concept of \textbf{Research Session}, which groups a sequence of consecutive research actions into a session (see the left column of Figure \ref{fig:context_management}).

Formally, a research session can be represented as:
\begin{equation}
s_k = \{a_i, a_{i+1}, ..., a_j\}, \quad \text{where } a_i, a_j \in \mathcal{A}_{\text{milestone}}
\end{equation}

where $s_k$ represents the $k$-th research session, and $\mathcal{A}_{\text{milestone}}$ denotes the set of milestone actions that mark session boundaries.

Each research session begins and ends with a milestone action. All \textbf{User Actions} and \textbf{Administrative Actions} are regarded as milestone actions. Additionally, the agent is required to execute a special \textbf{Process Information Action} to summarize current research progress when significant progress is made or when a preset number of action rounds is reached, and this action is also treated as a milestone action. This design allows users to quickly grasp the research process at a macro level by browsing through these milestone actions.

\subsection{Research Context Reduction}

Since existing LLM-based agents have context limits and their capabilities degrade as irrelevant information accumulates in the context \cite{hong2025context}, we have designed a research context reduction mechanism based on our hierarchical research context architecture to dynamically reduce context during the research process. A key insight is that during research, \textbf{Search Information} and \textbf{Source Information} are typically utilized within a short timeframe to support the generation of corresponding \textbf{Processed Information}, and the intermediate information generated within a \textbf{Research Session} is usually unimportant to other \textbf{Research Sessions}. Based on this insight, we immediately minimize previous \textbf{Search Information} and \textbf{Source Information} after the agent executes a \textbf{Process Information Action} (i.e., replacing the full text of the information with a pointer to that full text), and immediately minimize all information except the information generated by milestone actions (defined in Section \ref{Section:Level3: Research Session}) when starting a new \textbf{Research Session}. Through this mechanism, we effectively prevent the context from being filled up and exhausted by various types of information during long-horizon information seeking processes.

\subsection{Research Context Backtrace}
\label{Section:Research Context Backtrace}

To help users efficiently trace conclusions they care about back to their supporting evidence among numerous information blocks, we have designed a context backtrace mechanism based on our hierarchical research context architecture. Users can select any segment of text from any information generated during the research process and initiate context backtrace. The system then launchs a dedicated LLM-based agent to examine the prerequisite information identified through the \textbf{Action Dependency Graph} to determine whether it contains specific supporting evidence for the current information. If such evidence is found, the agent will record it and recursively examine the supporting evidence of these evidence pieces until tracing back to the original raw information.
\section{System Interface}

\begin{figure*}[t]
  \centering
  \includegraphics[width=\textwidth]{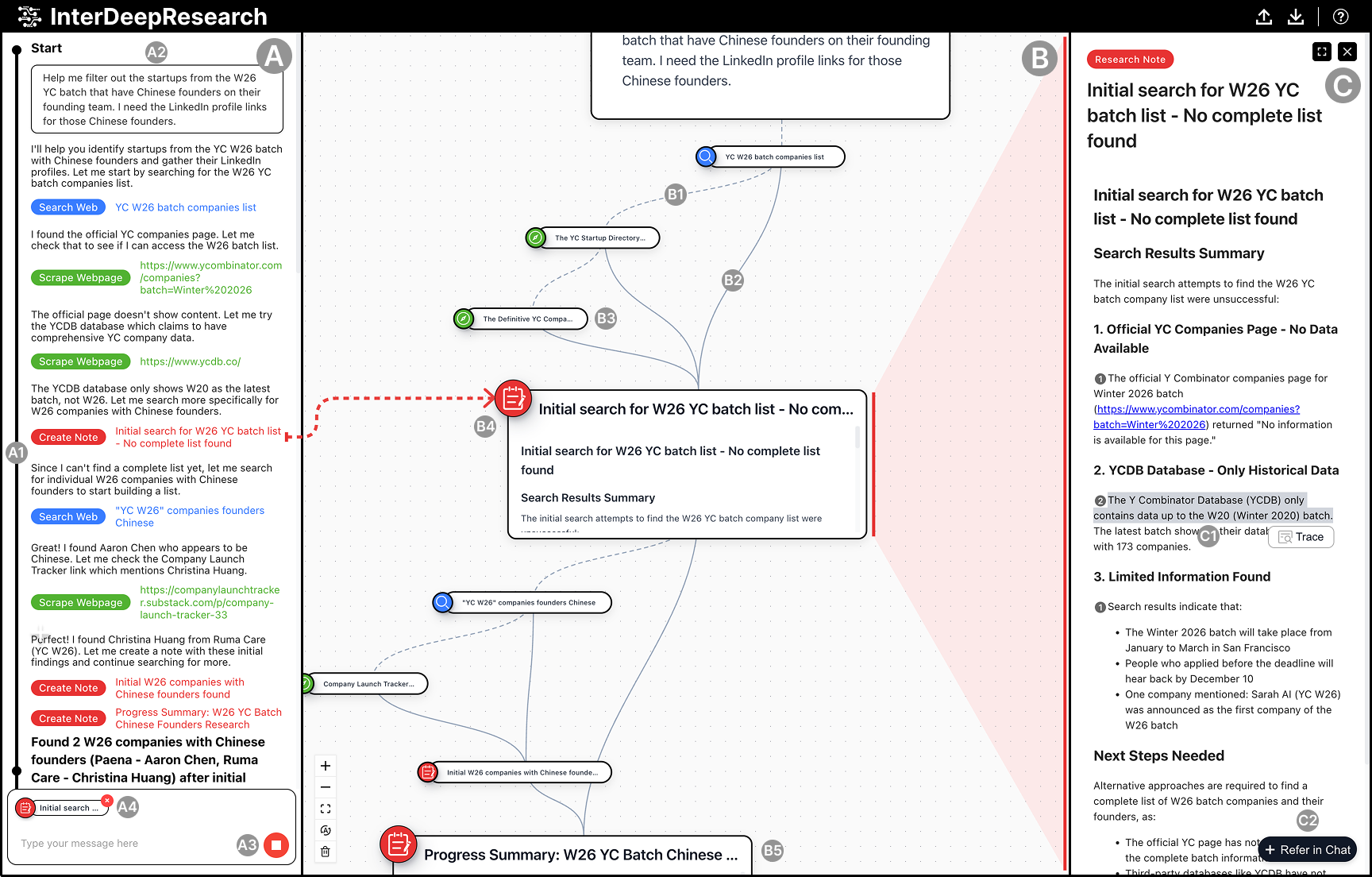}
  \vspace{-15pt}
  \caption{InterDeepResearch interface with three coordinated views: (A) chat-style view for research sessions and research action flow, (B) graph-based view for research action dependencies, and (C) card-style view for detailed research information.}
  \Description{A screenshot of the InterDeepResearch system interface showing three main panels. The left panel (A) displays a chat-like interface with colored action items and session markers. The center panel (B) shows a graph visualization with connected nodes representing research actions. The right panel (C) displays detailed information cards with text content.}
  \label{fig:system_interface}
  \vspace{-13pt}
\end{figure*}
In this section, we introduce the interface of \textit{interDeepResearch}.

\subsection{Visual Research Context Organization}
 To clearly present the agent's research process to users, we designed three coordinated views based on the hierarchical research context architecture described in Section \ref{section:Hierarchical Research Context Architecture}, providing an intuitive visual organization of the research context.

\subsubsection{View for Research Session \& Research Action Flow}
As shown in area \raisebox{-0.2em}{\includegraphics[height=1.0em]{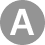}} of Figure \ref{fig:system_interface}, we employ a linear chat-style view to compactly display both the \textbf{Research Session} and \textbf{Research Action Flow}. Users can identify their current \textbf{Research Session} by checking the annotated line on the left side of the view (see Figure \ref{fig:system_interface}, \raisebox{-0.2em}{\includegraphics[height=1.0em]{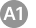}}), and understand the corresponding agent's research process by reviewing the \textbf{Research Action Flow} on the right side of the view (see Figure \ref{fig:system_interface}, \raisebox{-0.2em}{\includegraphics[height=1.0em]{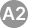}}). To improve understanding efficiency, we require the agent to provide concise narrative descriptions before and after each \textbf{Research Action} (i.e., what just happened and what it is about to do). We also employ color encoding to help users quickly identify the type of \textbf{Research Action} (i.e., blue for \textbf{Search Information Action}, green for \textbf{Source Information Action}, and red for \textbf{Processed Information Action}). During the research process, users can interrupt the agent's current \textbf{Research Action Flow} at any time by clicking the stop button at the bottom of the view (see Figure \ref{fig:system_interface}, \raisebox{-0.2em}{\includegraphics[height=1.0em]{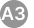}}) and sending arbitrary \textbf{User Information} to steer the current research direction. 

\subsubsection{View for Research Action Dependency Graph}
As shown in area \raisebox{-0.2em}{\includegraphics[height=1.0em]{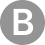}} of Figure \ref{fig:system_interface}, we employ a graph-based canvas-style view to display the \textbf{Research Action Dependency Graph}. Each node in the graph represents one \textbf{Research Action} and is arranged vertically from top to bottom according to execution order. We connect two consecutive \textbf{Research Actions} in the execution sequence with dashed lines (see Figure \ref{fig:system_interface}, \raisebox{-0.2em}{\includegraphics[height=1.0em]{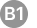}}), and connect two \textbf{Research Actions} that have information dependencies with solid lines (see Figure \ref{fig:system_interface}, \raisebox{-0.2em}{\includegraphics[height=1.0em]{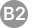}}). To increase information density and help users quickly locate key \textbf{Research Actions}, by default, all nodes are in a ``collapsed'' state (showing only key action parameters, see Figure \ref{fig:system_interface}, \raisebox{-0.2em}{\includegraphics[height=1.0em]{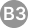}}) except for nodes that are currently focused (see Figure \ref{fig:system_interface}, \raisebox{-0.2em}{\includegraphics[height=1.0em]{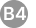}}), or corresponds to milestone actions (see Figure \ref{fig:system_interface}, \raisebox{-0.2em}{\includegraphics[height=1.0em]{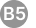}}), which are in an ``expanded'' state (displaying the information generated by that action). Users can toggle any node between ``expanded'' and ``collapsed'' states by right-clicking on it.
\subsubsection{View for Research Information}
As shown in area \raisebox{-0.2em}{\includegraphics[height=1.0em]{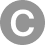}} of Figure \ref{fig:system_interface}, we employ a card-style view to display the \textbf{Research Information}. This view presents the detailed content of information generated during the research process. Users can click the ``\raisebox{-0.2em}{\includegraphics[height=1.0em]{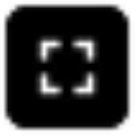}}'' icon to view the information in full screen. Additionally, since users may often need to reference \textbf{Information} generated during previous research processes when providing \textbf{User Information}, we allow users to select any portion they wish to reference (see Figure \ref{fig:system_interface}, \raisebox{-0.2em}{\includegraphics[height=1.0em]{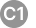}}) and click the ``Refer in Chat'' button (see Figure \ref{fig:system_interface}, \raisebox{-0.2em}{\includegraphics[height=1.0em]{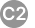}}) to reference it in their \textbf{User Information} (see Figure \ref{fig:system_interface} \raisebox{-0.2em}{\includegraphics[height=1.0em]{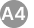}}).
\subsection{Interactive Research Context Navigation}
To support efficient research context navigation, we propose two dedicated interaction mechanisms, which are elaborated as follows.
\subsubsection{Cross-view Linkage}
To support users' dynamic navigation needs as they frequently switch between different views for cross-referential analysis, we enable users to ``focus'' on any \textbf{Research Action} in the interface by clicking on it, which quickly retrieves the corresponding information in each view and establishes visual linkage between them. For example, when a user clicks on a ``create note'' action in the \textbf{View for Research Session \& Research Action Flow}, the corresponding node in the \textbf{View for Research Action Dependency Graph} also expands and moves to the center of the canvas with a connection established (see Figure \ref{fig:system_interface}, \raisebox{-0.2em}{\includegraphics[height=1.0em]{figures/B4.png}}), while the \textbf{View for Research Information} simultaneously displays the corresponding \textbf{Information} card. Similarly, if a user clicks on a node in the \textbf{View for Research Action Dependency Graph}, the corresponding parts in other views also adjust and link to it. Additionally, when users click the ``\raisebox{-0.2em}{\includegraphics[height=1.0em]{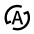}}'' icon, the system will dynamically ``focus'' on the most recently completed \textbf{Research Action}.

\subsubsection{Cross-action Backtrace}
\begin{figure}[h]
  \centering
  \includegraphics[width=\columnwidth]{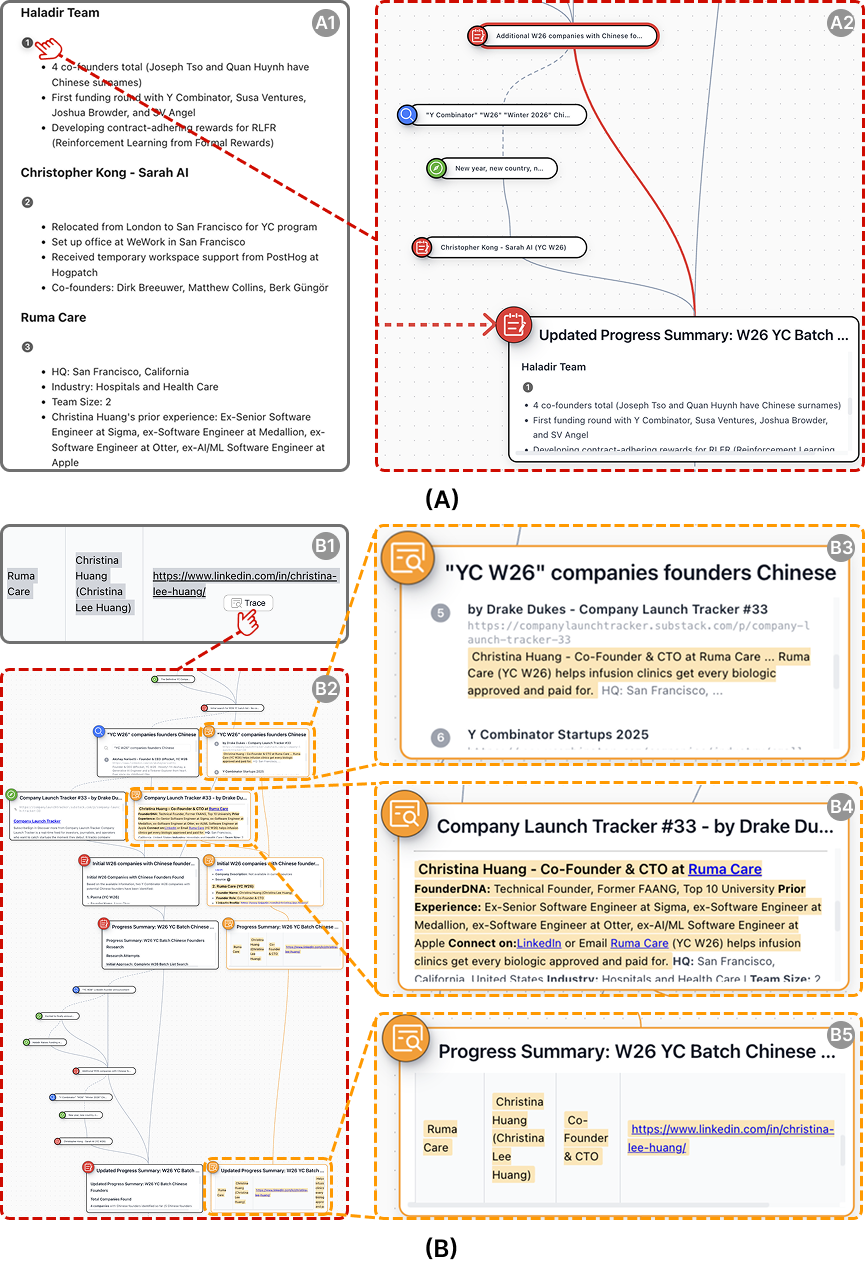}
  \vspace{-20pt}
  \caption{The cross-action backtrace mechanism helps users trace the provenance of generated content.}
  \Description{A diagram illustrating the backtrace mechanism with highlighted nodes and connections showing the path from a selected piece of information back through its dependencies to the original source information.}
  \label{fig:Backtrace}
  \vspace{-10pt}
\end{figure}

To help users trace the provenance of generated content and verify its supporting evidence, we require the agent to cite prior \textbf{Information} using a specific superscript format when generating \textbf{Processed Information}. When users hover their mouse over these superscripts (see Figure \ref{fig:Backtrace}, \raisebox{-0.2em}{\includegraphics[height=1.0em]{figures/A1.png}}), the corresponding nodes are automatically highlighted in the \textbf{View for Research Action Dependency Graph} (see Figure \ref{fig:Backtrace}, \raisebox{-0.2em}{\includegraphics[height=1.0em]{figures/A2.png}}). Additionally, users can proactively select a small segment of text in the \textbf{View for Research Information} and click the pop-up ``Trace'' button (see Figure \ref{fig:Backtrace}, \raisebox{-0.2em}{\includegraphics[height=1.0em]{figures/B1.png}}). The system then initiates the research context backtrace (Section \ref{Section:Research Context Backtrace}) to recursively examine supporting evidence in all prior dependent \textbf{Information}, and upon finding it, creates temporary nodes in the \textbf{View for Research Action Dependency Graph} to visualize the Backtrace results (see Figure \ref{fig:Backtrace}, \raisebox{-0.2em}{\includegraphics[height=1.0em]{figures/B2.png}}-\raisebox{-0.2em}{\includegraphics[height=1.0em]{figures/B5.png}}). Users can choose to retain these temporary nodes for later analysis or delete them by clicking the ``\raisebox{-0.2em}{\includegraphics[height=1.0em]{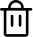}}'' button in the \textbf{View for Research Action Dependency Graph}.
\section{Usage Scenario}

To illustrate how \textit{InterDeepResearch} supports human-agent collaborative information seeking, we present a usage scenario featuring Alice, a headhunter researching Chinese founders in Y Combinator's Winter 2026 batch. \textbf{We strongly recommend watching the corresponding live demo on our project page.}

\textbf{Start \& Monitor:} Alice opens \textit{InterDeepResearch} and inputs: \textit{``Help me filter out the startups from the W26 YC batch that have Chinese founders on their founding team. I need the LinkedIn profile links for those Chinese founders.''} The agent begins working, displaying its research process in real-time through three coordinated views. After several minutes, Alice reviews the progress through milestone markers and focuses on the latest milestone action, finding the agent has identified five Chinese founders.

\textbf{Backtrace:}  Alice notices Christina Huang, a healthcare startup founder, and wants to verify its information source. Alice selects the text about Christina and clicks the ``Trace'' button. The system then performs a context backtrace, highlighting a trace path in the canvas view that reveals the information originates from a website named ``Company Launch Tracker'' and a search result page. 

\textbf{Change Strategy:} Looking at the trace results, Alice grows concerned that the agent is collecting information in a scattered manner without a systematic approach. As an experienced head hunter, she suspects some websites likely maintain complete lists of all W26 YC batch companies. Therefore, Alice clicks ``Stop'' to interrupt the agent and provides guidance: \textit{``Searching randomly like this isn't working. I think we should do our best to first try to find a complete list of all companies in the W26 YC batch, and then use that list to identify the Chinese founders.''} The agent adjusts its approach and soon discovers an open-source YC API providing the W26 company list in JSON format. It identifies that company slugs can be used to access YC profile pages containing founders' LinkedIn links, then proposes a systematic plan to process each company sequentially and starts executing.

\textbf{Correct Mistake:} As Alice examines the agent's recent progress, she notices the agent reporting ``404 file not found'' errors for many companies. She focuses on the webpage scraping actions for these companies to examine the errors. Alice suspects these company names weren't actually on the original list, so she navigates back to the YC W26 company list and, after checking, confirms that these names are indeed hallucinations by the agent rather than actual companies from the list. Alice immediately clicks the ``Stop'' button, selects the original company list, clicks ``Refer in Chat'' to quote it, and sends: \textit{``Please search strictly in the order of this list.''} The agent acknowledges its mistake and restarts.

\textbf{Result \& Continue:} After being corrected, the agent successfully processes each company, identifying Chinese founders and collecting their LinkedIn profiles. Upon completion, the agent provides a summary report containing all qualifying Chinese founders' information and LinkedIn links. The preserved research context allows Alice to dive deeper based on her new research intents, or transfer the accumulated experience to similar tasks such as researching other founders in different YC batches.


\section{Implementation \& Technical Evaluation}
\label{sec:Technical Evaluation}

We developed the system frontend using the React \cite{react2025} framework and implemented the draggable canvas interface with the Reactflow \cite{reactflow2025} library. Layout computation for visual elements was performed using D3 \cite{d3js2024} and ELKJS \cite{elkjs2025}. The backend was built in Python, using LiteLLM \cite{litellm2026} to standardize backend LLM API requests. Claude Sonnet 4.5 \cite{claude2025} served as the language model across all experiments.

\begin{figure}[t]
  \centering
  \includegraphics[width=\columnwidth]{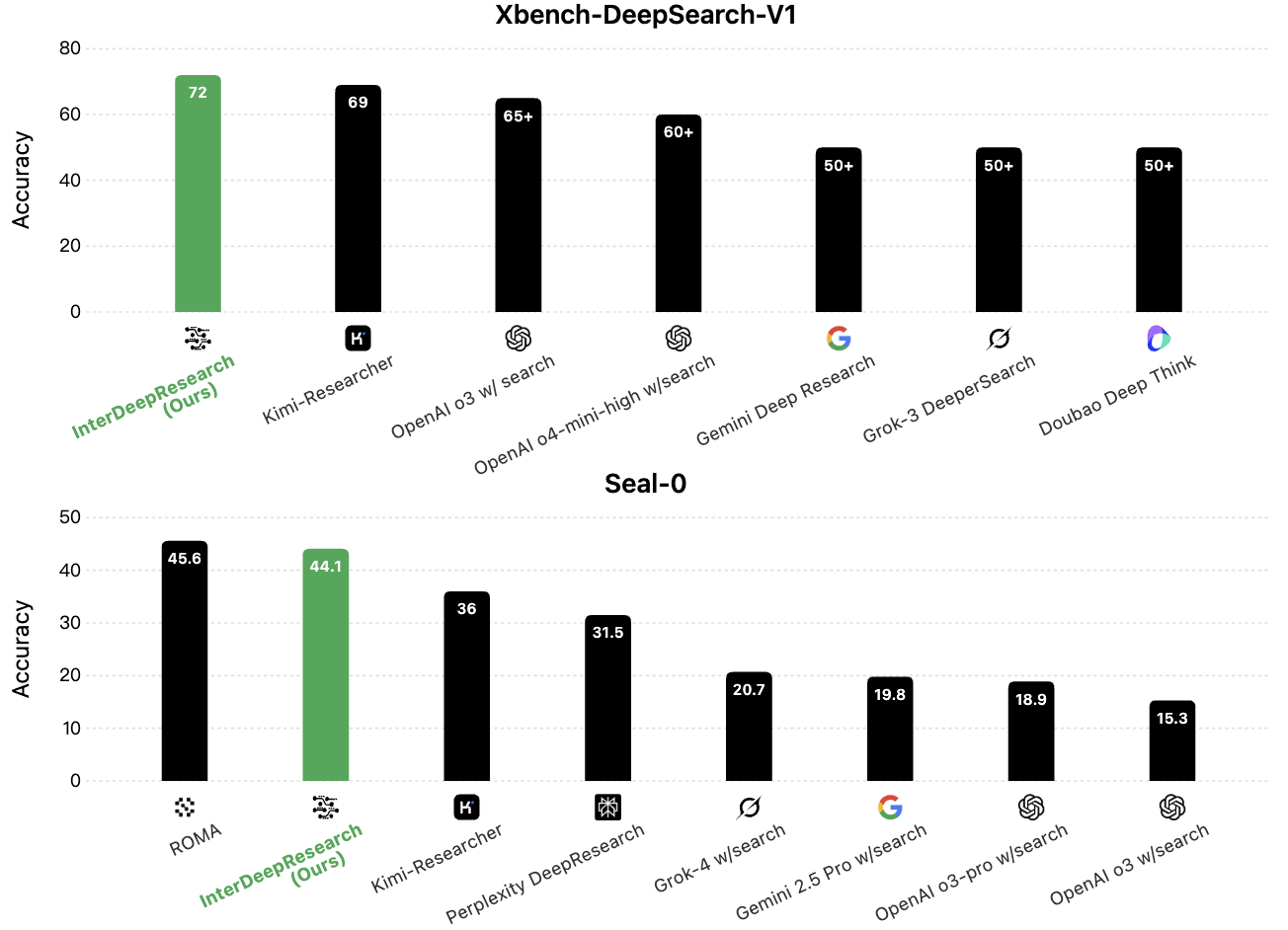}
  \vspace{-15pt}
  \caption{InterDeepResearch achieves competitive performance on existing text-based deep research benchmarks.}
  \Description{A bar chart or comparison table showing performance scores of InterDeepResearch against other deep research systems on two benchmarks: Xbench-DeepSearch-V1 and Seal-0. InterDeepResearch shows the highest scores compared to commercial systems.}
  \label{fig:techEval}
  \vspace{-15pt}
\end{figure}

To validate the basic information-seeking capabilities of \textit{InterDeepResearch}, we evaluate it (without user intervention) on two representative text-based deep research benchmarks (end-to-end execution without any user intervention): Xbench-DeepSearch-v1\cite{xbench} and Seal-0\cite{sealqa}. We then compared \textit{InterDeepResearch}'s score against the top-performing systems on the benchmark leaderboards (see Fig. \ref{fig:techEval}). The results demonstrate that InterDeepResearch achieves competitive performance in automated information seeking, even surpassing popular commercial deep research systems like Perplexity Deep Research \cite{perplexity_deep_research} and Gemini Deep Research \cite{gemini_deep_research}.

\section{User Study}
\label{sec:User Study}
The primary goal of this study is to evaluate the effectiveness of \textit{InterDeepResearch} in facilitating human-agent collaborative information seeking. To assess the efficacy of our system, we conducted a user study with 15 participants.

\subsection{Participants}
We recruited 15 frequent users (U1-U15) who extensively utilize existing deep research systems for our user study. Among them, 4 had experience developing deep research systems, 12 regularly used them in their work, and 7 in their daily lives.
\subsection{Procedure}

\textbf{Introduction and Training:} We began by providing participants with a brief introduction to the study's background and objectives, then proceeded with a guided tour of all features of \textit{InterDeepResearch}. Participants were then given ample time to explore and familiarize themselves with the system. We encouraged participants to ask questions at any point during this process.

\textbf{Task Process:} 
Participants were asked to first select from our provided task pool (covering four different themes: academic, business, development, and entertainment) an information-seeking problem they were most interested in exploring collaboratively with the agent (e.g., ``Find as many games as possible on the Steam platform that have AI as their core gameplay mechanic. Need to provide the game name, developer, AI application scenario, release date, and Steam link.''). They would then use \textit{InterDeepResearch} to explore and address this task. After completing the task, we invited users to propose an information-seeking task of personal interest for open-ended exploration. Throughout the process, we asked participants to think aloud, continuously sharing interesting discoveries and their feedback about the system.



\begin{figure}[t]
  \centering
  \includegraphics[width=\columnwidth]{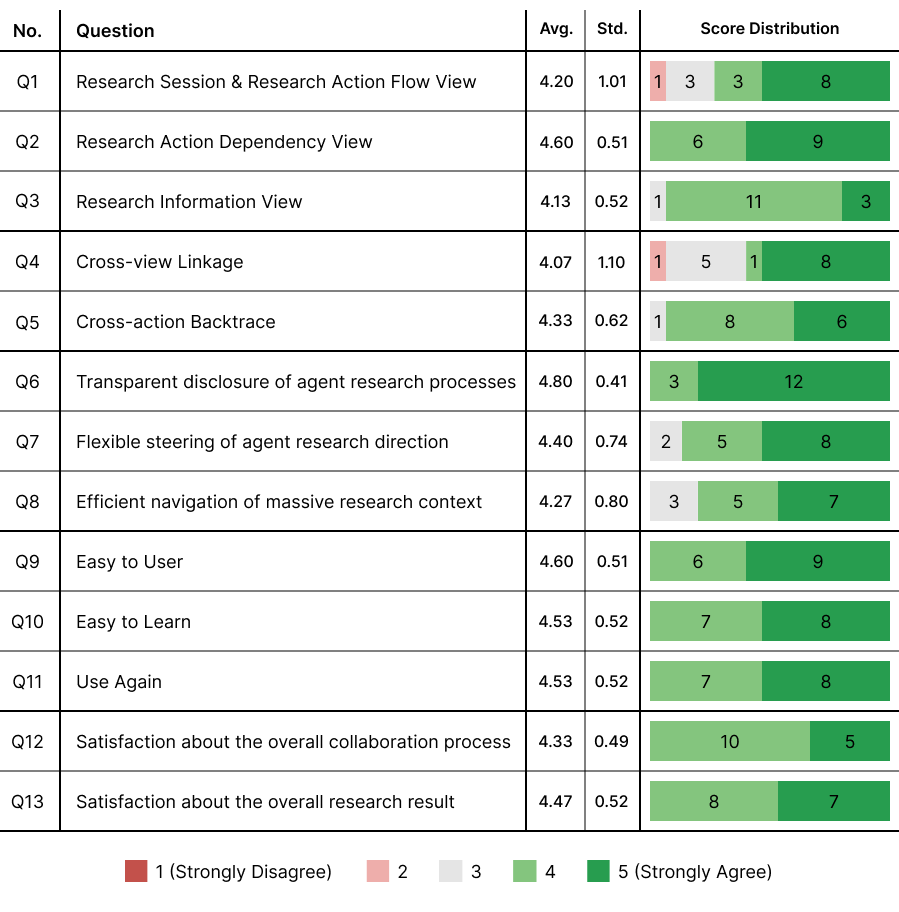}
  \vspace{-20pt}
  \caption{The results of the questionnaire regarding our system’s effectiveness and usability.}
  \Description{A visualization showing user study results with Likert-scale ratings for different aspects of the system including system views effectiveness, interaction mechanisms, support for collaborative information seeking, system usability, and overall satisfaction. Most ratings are above 4.0 on a 5-point scale.}
  \label{fig:userStudy}
  \vspace{-15pt}
\end{figure}

\textbf{Semi-structured Interview:} After completing the task, participants were asked to complete a five-point Likert-scale questionnaire to evaluate the effectiveness of our system's views and interactions, and its support for human-agent collaborative information seeking (Figure \ref{fig:userStudy}). For each question, we encouraged participants to explain their ratings and share any additional thoughts. We also encouraged them to compare their current experience with \textit{InterDeepResearch} to their prior experience with their favorite deep research systems. 

\subsection{Result Analysis}

\subsubsection{Effectiveness of System Views}
Overall, all views received positive feedback from most users for their effectiveness in facilitating collaborative information seeking with the agent. The \textbf{Research Action Dependency Graph View}, in particular, received overwhelmingly positive feedback ($\mu = 4.7, \sigma = 0.5$). Multiple users mentioned that this graph-based representation was ``more intuitive compared to the linear text representation in the left-side view'' (U1, U2, U5, U6), ``helpful for information tracing'' (U1, U2, U6, U9, U11, U12), ``facilitated intuitive understanding of the Agent's research process'' (U2, U6, U10, U14), and ``significantly increased confidence'' (U4, U9). The \textbf{Research Information View} was generally well-received by most users ($\mu = 4.13, \sigma = 0.52$), who commonly found its design reasonable and frequently useful. An interesting finding was that user ratings of the \textbf{Research Session \& Research Action Flow View} showed divergence ($\mu = 4.20, \sigma = 1.01$): some users highly valued this view's role during the research process, frequently using it to ``quickly track research progress'' (U1, U2, U8, U14) and ``understand the agent's behavior'' (U2, U11, U15); while other users indicated they ``used this view sparingly, preferring to rely more heavily on the research action dependency view'' (U4, U6, U9, U10).

\subsubsection{Effectiveness of System Interactions}
The cross-view linkage mechanism received mixed feedback ($\mu = 4.07, \sigma = 1.10$): some users greatly appreciated this mechanism, finding it highly intuitive and effective in improving the efficiency of locating information during lengthy research processes (U1, U2, U7, U8, U11, U14, U15); while other users indicated they did not benefit much from this interaction mechanism (U3, U4, U6). This result echoes the divergence in user attitudes toward the \textbf{Research Session \& Research Action Flow View}, revealing two distinct usage patterns: one usage pattern involves users constantly switching between different views based on their current navigation needs—sometimes navigating via the \textbf{Research Action Flow}, sometimes via the \textbf{Research Action Dependency Graph}, and sometimes wanting to cross-reference both views for analysis; the other usage pattern involves users continuously focusing on the \textbf{Research Action Dependency Graph}, performing the vast majority of navigation and analysis within that view. For the former, the cross-view linkage mechanism proved highly useful, significantly improving their analysis efficiency and coherence, while for the latter, this mechanism was of limited utility. The cross-action backtrace mechanism was appreciated by nearly all users ($\mu = 4.33, \sigma = 0.62$), who found it greatly improved the efficiency of information tracing and frequently used it when wanting to verify important information or when encountering unexpected conclusions. However, some users also expressed that the current system requires relatively long wait times after clicking ``trace,'' and hoped for optimized wait times (U2, U3, U14).

\subsubsection{Support for Human-Agent Collaborative Information Seeking}
All users agreed that our system effectively supported clear comprehension of agent research processes ($\mu = 4.80, \sigma = 0.41$). U4 described our system as ``the only deep research system I've seen so far that truly allows people to follow the system's research process'' (U4). Most users also found that our system adequately supported flexible steering of agent research direction ($\mu = 4.40, \sigma = 0.74$). U3 told us, ``With this system, adjusting the research direction is as easy as with a chat system. I just need to tell it what I want and it follows well, unlike Gemini Deep Research where I have to wait dozens of minutes each time for it to finish running before making requests, and it doesn't necessarily follow my requirements'' (U3). However, multiple users also suggested that the system could provide more mechanisms for steering the agent. For example, U2 proposed supporting the deletion of specific prior context (U2), and U5 suggested supporting direct editing of previous information (U5). Most users also felt that our system adequately supported efficient navigation of research context ($\mu = 4.27, \sigma = 0.80$). U11 told us, ``Compared to previous deep research systems where you could only manually scroll through the final long text and click through cited source links one by one to search for information, this is a significant improvement in both efficiency and experience.'' Some users also proposed further optimization of the navigation experience, such as providing more information simplification and management support as research steps accumulate (U6, U8), and allowing users to jump back with one click after navigating to a previous action through trace mechanisms (U10).

\subsubsection{System Usability}

All users found our system to be ``easy to use'' ($\mu = 4.60, \sigma = 0.51$) and ``easy to learn'' ($\mu = 4.53, \sigma = 0.52$). Users also indicated they would like to use our system again in the future ($\mu = 4.53, \sigma = 0.52$). Multiple users mentioned that for future research tasks requiring high reliability and customizability, they would significantly prefer using our system over other deep research systems (U1, U2, U4, U15).

\subsubsection{Overall Satisfaction}


All users expressed high satisfaction with the collaborative process of working with the agent using our system ($\mu = 4.33, \sigma = 0.49$). U6 told us, ``In traditional systems, humans are just passive. You can only give high-level requirements, and if the final result doesn't meet expectations, you have to start over and guide it again, and the agent doesn't necessarily listen to your guidance during execution. But with this system, I can see the information flow throughout the entire research process, and I can intervene as soon as I spot an issue. The collaboration efficiency is significantly higher, and it has greatly increased my confidence in the collaboration.'' All users also expressed high satisfaction with the overall research results ($\mu = 4.47, \sigma = 0.52$). U1 praised that ``both the research process and the final result feel highly credible. It's a result under my control rather than randomly opening a mystery box'' (U1). U4 told us, ``Because I gradually built trust with the agent during our collaboration, I can be quite confident that the final research report contains truthful information. Unlike systems like Gemini Deep Research where the content is rather mixed and I don't feel confident trusting it'' (U4). Furthermore, some users indeed discovered high-value results with real-world significance while using our system. For example, U6 used our system to collect job opportunities across the web, and through multiple rounds of collaboration with the agent, continuously optimized the information gathering strategy and found many valuable job leads. When tracing the source of one piece of information he was particularly interested in, he discovered it came from an information source he had never imagined (a niche study abroad forum), and found even more high-value information for himself on that forum.

\section{Conclusion}
In this work, we presented InterDeepResearch, the first interactive deep research system that enables human-agent collaborative information seeking. Our system transforms users from passive recipients into active collaborators who can clearly comprehend agent research processes, flexibly steer research directions, and efficiently navigate massive research context, taking an early step toward unlocking the potential of human-agent synergy in complex information seeking tasks.

While our results offer a promising early signal for the value and viability of human-agent symbiosis for information seeking, several avenues remain for future advancement. For instance, our user study revealed that different users may exhibit distinct preferences and usage patterns across different scenarios, suggesting the value of more adaptive interfaces during collaboration. Moreover, it is worth exploring how richer modalities such as hybrid input (e.g. voice-based) and multimodal information presentation could make the research process more efficient and engaging. Extending beyond single-agent-single-user settings to support multiple agents, multiple users, and branching research threads could enable higher information processing bandwidth and more comprehensive research coverage, though the design challenges in such complex collaborative scenarios remain interesting open questions.



\bibliographystyle{ACM-Reference-Format}
\bibliography{template}

\end{document}